\documentclass{article}
\usepackage{amssymb}
\usepackage{amsmath}
\usepackage{epsfig}

\newcommand{\bea}{\begin{eqnarray}} 
\newcommand{\eea}{\end{eqnarray}}

\newcommand{\slB}{\raise.15ex\hbox{$/$}\kern-.53em\hbox{$B$}}
\newcommand{\slA}{\raise.15ex\hbox{$/$}\kern-.7em\hbox{$A$}}
\newcommand{\slL}{\raise.15ex\hbox{$/$}\kern-.53em\hbox{$L$}}
\newcommand{\slD}{\raise.15ex\hbox{$/$}\kern-.7em\hbox{$D$}}
\newcommand{\slP}{\raise.15ex\hbox{$/$}\kern-.63em\hbox{$P$}}
\newcommand{\slp}{\raise.15ex\hbox{$/$}\kern-.53em\hbox{$p$}}
\newcommand{\slk}{\raise.15ex\hbox{$/$}\kern-.53em\hbox{$k$}}
\newcommand{\slq}{\raise.15ex\hbox{$/$}\kern-.53em\hbox{$q$}}
\newcommand{\slR}{\raise.15ex\hbox{$/$}\kern-.53em\hbox{$R$}}
\newcommand{\slQ}{\raise.15ex\hbox{$/$}\kern-.63em\hbox{$Q$}}
\newcommand{\slK}{\raise.15ex\hbox{$/$}\kern-.7em\hbox{$K$}}
\newcommand{\slSigma}{\raise.15ex\hbox{$/$}\kern-.53em\hbox{$\Sigma$}}
\newcommand{\slcalP}{\raise.15ex\hbox{$/$}\kern-.63em\hbox{$\cal P$}}
\numberwithin{equation}{section}
\begin{document}
\pagestyle{myheadings}  \markright{Version of \today  }

\title{Dileptons from hot heavy static photons}

\author{P. Aurenche\footnote{Electronic address: {\tt aurenche@lapp.in2p3.fr}}\\ Laboratoire d'Annecy-le-Vieux de Physique Th\'{e}orique, 74941 Annecy-le-Vieux, France \and M. E. Carrington\footnote{Electronic address:
{\tt carrington@brandonu.ca}} \\ Department of Physics, Brandon University, Manitoba, R7A 6A9 Canada \and
N. Marchal \\ Laboratoire d'Annecy-le-Vieux de Physique Th\'{e}orique, 74941 Annecy-le-Vieux, France}
\maketitle

\begin{abstract}
We compute the production rate of lepton pair by  static photons at finite
temperature at two-loop order. We treat the infrared region of the gluon phase space carefully by using a hard thermal loop gluon propagator.  The
result is free of infrared and collinear divergences and exhibits an enhancement which produces a result of order $\sim e^2 g^3$ instead of $\sim e^2 g^4$ as would be expected from ordinary perturbation theory.
\end{abstract}

\section{Introduction}

The production of photons and leptons at finite temperature is studied as a signal for
the formation of a quark-gluon plasma in heavy ion collisions~\cite{Peitzmann:2001mz,Aggarwal:2000th,Aggarwal:2000ps,Srivastava:ct,Srivastava:2000pv,Huovinen:2001wx,Sollfrank:1996hd}. Photons and leptons are expected to escape easily from nuclear matter
because of their large mean free path. Computations are done in two steps. The first step is to compute the production rate at a given order in perturbative thermal field theory. The second step is to substitute this result into a  hydrodynamical model that simulates the experimentally realized heavy ion collision~\cite{Srivastava:ct,Srivastava:2000pv,Huovinen:2001wx,Sollfrank:1996hd}. In this paper we discuss only the first of these steps.

The production rate of lepton pairs by a photon of
virtuality~$Q^2$ is~\cite{GaleKapusta91}

\begin{equation}\label{prodrate}
  \frac{dN}{dq_0 \, d^3 q \, d^4 x} = \frac{\alpha}{12 \pi^3}
  \frac{1}{Q^2} n_B (q_0) \, \text{Im}\, \Pi_R(q_0,q)^{\mu}_{~\mu}
\end{equation}
Note that there are two coupling constants involved in this
equation, or equivalently, two fine structure constants: $e$ or
$\alpha$ characterizes the electromagnetic interaction, and $g$ or
$\alpha_s$ characterizes the strong interaction. The formula above
is valid to first order in $\alpha$ and arbitrary order in
$\alpha_s$, although the actual calculation will involve a
perturbative expansion of the polarization tensor.

In this paper we are interested in contributions to lepton pair
production by heavy static photons. We take the photon momentum to
be: $Q^{\mu }= (q_0, \vec{0})$ with $Q^2=q_0^2\gg g^2T^2$. At the Born
level, the imaginary part of the trace of the polarization tensor
is given by~:
\begin{equation}\label{Born}
  \text{Im} \, \Pi_R(q_0,\vec{0})^{\mu}_{~\mu} =  \alpha \ N_c \ q_0^2 \left[
1-2 \, n_F \left( \frac{q_0}{2} \right) \right],
\end{equation}
which we denote, for simplicity, $\text{Im}\, \Pi^{\rm Born}$. $N_c$ is the
number of colors.

The rate of static lepton pair production has also been calculated at the
two-loop level in the bare theory. In physical terms these corrections include
the contributions of Compton scattering, quark-antiquark annihilation into a 
virtual photon, and the 3 $\rightarrow$ 1 process $q \bar q G \rightarrow
\gamma^*$. In the large $q_0^2/T^2$ limit, the result has been found to 
be~\cite{Altherr:1989jc}:
\begin{equation}
\label{eq:2-bareloop}
\text{Im} \, \Pi_R(q_0,\vec{0})^{\mu}_{~\mu} =  \text{Im}\, \Pi^{\rm Born}
\, \sqrt{1 - {4 m^2_{\rm q} \over q_0^2}} \, 
\left[ 1 + {2 m^2_{\rm q} \over q_0^2}  +  C_F\ {3 \over 4} {g^2 \over 4 \pi^2} \right]
\end{equation}
where $m^2_{\rm q} = C_F g^2\ T^2 / 4$ is the thermal quark mass generated by
the one-loop self-energy corrections, with $C_F =  4/3$ the usual QCD
structure constant. The temperature independent term is simply the $T=0$ first
order QCD correction to the process $e^{+} e^{-} \rightarrow q \bar q$ with massless kinematics. The
square root threshold factor in eq.~(\ref{eq:2-bareloop}) is obtained if the
thermal self-energy corrections are resummed on the quark lines. The
interpretation of the temperature dependent corrections is very simple.
Starting from massless quarks the thermal corrections, in the considered
limit, are obtained by simply considering the production in the vacuum of
quarks with a thermal mass. This result supports the quasi-particle
interpretation of thermal effects. In the strict perturbative sense, which we
adhere to here, the threshold factor can be expanded in $m^2_{\rm q}/q_0^2$
and one observes a cancellation of thermal corrections at ${\cal O}(g^2)$: the
first thermal corrections are then expected to appear ${\cal O}(g^4)$,
\begin{equation}
\text{Im} \, \Pi_R(q_0,\vec{0})^{\mu}_{~\mu} =  \text{Im}\, \Pi^{\rm Born}
\, \left[ 1 +  C_F\ {3 \over 4} {g^2 \over 4 \pi^2} + {\cal O}(g^4 {T^4\over
q_0^4}) \right]
\label{eq:2-bareloop2}
\end{equation}

These calculations are done by integrating over the full fermion and gluon
phase space including contributions from both hard and soft modes. However, it
is well known that it is inconsistent to use bare propagators at soft momentum
scales $(p\sim gT)$. For soft momenta, collective modes play an  important role
and hard thermal loop (HTL) resummed propagators must be used
\cite{Pisarski:cs,Braaten:1989kk,Frenkel:br}. We expect that this problem is
more serious for gluon modes than for fermion modes. Gluons are bosons and
carry a thermal factor of the Bose-Einstein form which has the limiting form
$T/l_0 \sim 1/g \gg 1 $ when $l_0$ (the gluon energy) is $\sim gT$. In
contrast, fermions carry a thermal factor of the Fermi-Dirac form which
approachs $1/2$ in the high temperature limit. Thus we expect that collective
effects due to soft gluons give rise to an enhancement.

\begin{figure}[ht]
\centering
\includegraphics[width=0.6\textwidth]{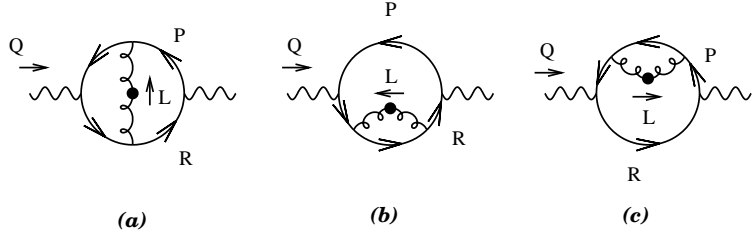}
\caption{Vertex {\it (a)} and self-energy {\it ((b) and (c))} diagrams}
\label{figdiagrams}
\end{figure}

In this article we will calculate the two loop photon polarization tensor using
a HTL gluon propagator in order to study the importance of the soft part of the
gluon phase space (Fig.~[\ref{figdiagrams}]). When the resummed gluon
propagator is used the physics is changed in the following ways:  

\noindent [1] transverse and longitudinal modes have different behaviour and
must be considered separately;

\noindent [2] in the time-like region gluons become massive
quasi-particles;

\noindent [3] in the space-like region the gluon polarization tensor has a
non-zero imaginary part which leads to Landau damping. This damping is
associated with new physical processes which do not appear at the bare two loop
level and correspond to the scattering of quarks and gluons in the medium.

One interesting aspect of the study of photon and lepton pair production at
finite temperature is the treatment of infrared and collinear divergences.  At
zero temperature there is a general result (the KLN
theorem~\cite{Kinoshita:ur,Lee:is}) which insures the cancellation of these
divergences.  At finite temperature, the problem is complicated by the presence
of the collective modes. In all calculations  divergences are removed through
cancellations that are similar to those that occur at zero temperature,
however, there is no generalized version of the KLN theorem that applies at
finite temperature.  

In bare perturbation theory, it has been shown that the corresponding diagrams 
are finite with respect to collinear and infrared divergences~\cite{Baier:1988xv,Altherr:1988bg,Altherr:1989yn,Gabellini:1989yk,Gale:2001}.
The case of a non-static photon has been studied
in~\cite{Cleymans:em,Wong:1993}. A calculation using resummed propagators has
been done in \cite{LeBellac:qg}. The authors use a scalar external particle,
bare fermion propagators and a HTL gluon propagator. They find  that all
divergences cancel, and that the contribution from longitudinal gluon modes is
enhanced relative to the contribution from transverse gluon modes by one power
of the coupling constant. The transverse mode gives a contribution to the
polarization tensor of order $e^2 {\cal O}(g^4)$, as is expected from
perturbation theory, and the longitudinal mode gives a contribution of order
$e^2{\cal O}(g^3)$. In this paper, we perform the corresponding calculation for
a virtual photon  (for which the result is related to the dilepton production
rate (\ref{prodrate}) which is a physical quantity). We find that the
cancellation of divergences has the same structure as in the scalar theory, but
we obtain an enhanced result $\sim e^2 g^3$ in both the longitudinal and
transverse sectors.   

This paper is organised as follows~: in Sec.~\ref{HTL} we review some elements
of real time finite temperature field theory and of the hard thermal loop
effective theory. In Sec.~\ref{Calculation} we outline the calculation of the
2-loop diagrams. We discuss some preliminary results in  Sec.~\ref{Results}. In
Sec.~\ref{Int} we perform the final integrals over the gluon energy and
momentum. Finally, in Sec.~\ref{Conclusions} we present some conclusions.

\section{Hard thermal loops}

\label{HTL}

We consider the case of hard fermions and soft gluons. Fermion propagators have
momenta of order~$p\gg gT$, and are not modified by collective
effects.  The fermionic propagator is given by
$$
S(P) = \slP D (P)\, ;~~D(P)= \frac{i}{P^2} = \frac{i}{p_0^2-\vec{p}^2}
$$
Retarded and advanced functions are obtained by taking the appropriate 
$i \epsilon$ prescription
$$
D(P)\rightarrow D_{R/A}(P) = \frac{i}{(p_0\pm i\epsilon)^2-\vec{p}^2}
$$
We also define the spectral function and the principal part,
\begin{subequations}
\label{rho}
\begin{align}
& \rho(P) = D_R (P) - D_A (P)\,= 2\pi\,\, \epsilon(p_0)\,\, \delta(P^2) = -\frac{\pi}{p} \sum_{n = \pm 1} n \,\,\delta(p_0+n p)  \\
&{\cal P}(P) = \frac{1}{2}\left(\frac{1}{(p_0+i\epsilon)^2-p^2}+\frac{1}{(p_0-i\epsilon)^2-p^2}\right)
\end{align}
\end{subequations}
which gives,
\begin{subequations}
\label{convert}
\begin{align}
& D_R(P) = i{\cal P}(P) + \frac{1}{2}\rho(P) \\
& D_A(P) = i{\cal P}(P) - \frac{1}{2}\rho(P) 
\end{align}
\end{subequations}

The exchanged gluon is soft, and thus the propagator will be modified by collective effects. We  use a hard thermal loop effective gluon propagator. The gluon propagator is diagonal in
color, and color indices will be omitted.
We work in the strict Coulomb gauge and decompose the gluon propagator into transverse and
longitudinal pieces:
\begin{equation}\label{DecompTLG}
  -D^{\mu \nu}_{R/A} (L) = P_T^{\mu \nu} (L) \; \Delta^T_{R/A} (L)
 + P_L^{\mu \nu} (L) \; \Delta^L_{R/A} (L)
\end{equation}
The projectors are defined as~\cite{Landsman:uw}~:
\begin{subequations}
\label{projectors}
\begin{align}
  P_T^{\mu \nu}(L)  & = \left( \begin{array}{cc} 0 & 0\\
  0 & - \delta^{ij}+ \hat{l}^i \hat{l}^j
  \end{array}
  \right)\\
  P_L^{\mu \nu}(L)  & = -\frac{L^2}{l^2}
  \left( \begin{array}{cc}
  1 & 0 \\ 0 & 0
  \end{array}\right)
\end{align}
\end{subequations}
and coefficients are:
\begin{align}
\label{DeltaTL}
 \Delta^{T/L} (L) & = \frac{i}{L^2 - \Pi_{T/L} (L)}
\end{align}
The quantities $\Pi_{T/L} (L)$ are defined by choosing the decomposition of the
gluon polarization tensor that satisfies the usual Schwinger-Dyson equation:
\begin{eqnarray}
\label{SDeqn}
&& \Pi^{\mu\nu}(L) = P_T^{\mu \nu}(L)\Pi_T(L) + P_T^{\mu \nu}(L)\Pi_T(L)  = i((D^{\mu\nu}(L))^{-1} - (D_0^{\mu\nu}(L))^{-1})
\end{eqnarray}
where $D_0^{\mu\nu}(L)$ is obtained from (\ref{DecompTLG}) by setting  the polarization tensor to zero. We note that (\ref{SDeqn}) holds separately for retarded and advanced components. The subscripts $R/A$ have been omitted to simplify the notation.
We use the 1-loop HTL results
~\cite{Weldon:aq,Klimov:bv} :
\begin{subequations}\label{PiTL}
\begin{align}
\label{PiT}
  \Pi_T (lx, l) & = 3\,  m_g^2 \left[ \frac{x^2}{2} + \frac{x (1-x^2)}{4} \ln \left( \frac{x+1}{x-1} \right) \right] \\
\label{PiL}
  \Pi_L (lx, l)  & = 3\,  m_g^2 \,  (1- x^2) \left[ 1 -
  \frac{x}{2} \ln \left( \frac{x+1}{x-1} \right) \right]
\end{align}
\end{subequations}
where~$x=l_0 /l$ and~$m_g^2 = g^2 T^2 [ N +N_f/2] /9$ is the soft gluon
thermal mass in a~$SU (N)$ gauge theory with~$N_f$ flavors. The spectral function is defined as
\bea
\label{spectral}
\rho_{T/L} (l_0,l) = \Delta^{T/L}_{R}
(L)- \Delta^{T/L}_{A} (L) .
\eea

\section{Calculation}
\label{Calculation}

We want to calculate the trace of the imaginary part of the retarded
photon polarization tensor which is given by
$$
{\rm Im} (\Pi_R)^{\mu}_{~\mu} =\frac{1}{2i}\left(
(\Pi_R)^{\mu}_{~\mu}   -  (\Pi_A)^{\mu}_{~\mu}
\right)
$$
The diagrams to be evaluated are shown in Fig. [1]. Since we consider the production of a static external photon
($Q^{\mu} = (q_0 , \vec{0})$), with $q_0>0$, all the azimuthal angular
integrations can be carried out trivially and the integration measure is
rewritten as,
\begin{equation}
  \int d^4 P \int d^4 L = 8 \pi^2 \int
  dl_0\int dl \ l^2 \int d y \int dp_0 \int dp \ p^2 
\label{phasespace}
\end{equation}
where $y = \hat l \cdot \hat p$ is related to the angle between the gluon and
quark momenta. We simplify the notation by defining the momentum variable
$R=P+Q$.

\subsection{Vertex diagram}

The vertex contribution (Fig. [1a]) to the imaginary part of the photon
polarization tensor can be written as~\cite{Aurenche:1991hi,Aurenche:1993vt,Aurenche:1996sh}:
\bea
\text{Im} {\Pi_R^{(vertex)}}^{\mu}_{\mu} &=&  - N_c C_F {e^2 g^2 \over 2} \int 
\frac{d^4 P}{ ( 2\pi )^4} \left[ n_F (r_0) - n_F (p_0) \right]  \int \frac{d^4
L}{ ( 2 \pi )^4}\, \text{Tr}_{\rho \sigma}^{(vertex)} n_B (l_0) {\cal
R}^{\rho\sigma}  
\nonumber \\
&&\qquad\qquad\qquad\times \ \Bigl[D_R(P) D_R(P+L) - D_A(P) D_A(P+L) \Bigr] \nonumber \\
&&\qquad\qquad\qquad\times \ \Bigl[D_R(R) D_R(R+L) - D_A(R) D_A(R+L) \Bigr]
 \label{vertex2-2}
\eea
where we have kept only terms proportional to the gluon thermal
factor $n_B(l_0)$ since we are interested in a possible enhancement from the soft part of the
gluon phase space.
Rewriting this expression using (\ref{convert}) we obtain,
\bea
\text{Im} {\Pi_R^{(vertex)}}^{\mu}_{\mu} &=&  N_c C_F e^2 g^2  \int 
\frac{d^4 P}{ ( 2\pi )^4} \left[ n_F (r_0) - n_F (p_0) \right]\;\rho(P)  \int
\frac{d^4 L}{ ( 2 \pi )^4}\, \text{Tr}_{\rho \sigma}^{(vertex)} n_B (l_0) {\cal
R}^{\rho\sigma}  
\nonumber \\
&& \qquad\qquad\qquad\times
~~{\cal P}(L+P)~\Bigl[ \rho(R+L) {\cal P}(R)  + \rho(R) {\cal P}(R+L) \Bigr]
 \label{vertex2}
\eea
where we have defined
\bea
{\cal R}^{\rho\sigma} = \rho_T  (L)\;  P_T^{\rho \sigma} (L) +  \rho_L  (L) \;
P_L^{\rho \sigma} (L)\; , \nonumber
\eea
and the Dirac trace is calculated to be: 
\bea
\text{Tr}_{\rho \sigma}^{(vertex)}  =&& \text{Tr} \left[ \slP \
\gamma^{\mu} \ \slR \ \gamma_{\rho} \ ( \slR + \slL ) \
\gamma_{\mu} \ ( \slP + \slL ) \ \gamma_{\sigma}\right]  \\
=&& - 8(2L\cdot P L\cdot R +L\cdot R P^2 +L\cdot P R^2+P^2R^2-L^2 R\cdot P)
g_{\rho\sigma}\nonumber \\ 
&& + 16 (R^2(L_\rho+P_\rho)P_\sigma +P^2(L_\rho+R_\rho)R_\sigma +L\cdot R
P_\rho(L_\sigma+P_\sigma-R_\sigma)\nonumber \\ &&+L\cdot P
R_\rho(L_\sigma+R_\sigma-P_\sigma)-L^2P_\rho R_\sigma-R\cdot P(P_\rho
R_\sigma+(L_\rho+P_\rho)(L_\sigma+R_\sigma))\; .\nonumber
\eea
By analogy with the zero temperature calculation, the first
term in eq.~(\ref{vertex2}) is called the real cut (the fermion propagators $P$
and $R+L$ are cut) while the second term  is called the virtual cut ($P$ and
$R$ are cut). There are two additional cuts: one where the fermion propagators carrying momenta  
$P+L$ are $R$ are cut, and one where the propagators carrying momenta $P+L$ and $R+L$ are cut. Using symmetry arguments one can show that these two terms contribute an overall factor of two which has been included in (\ref{vertex2}). 
The contraction $P_{T/L}^{\rho\sigma} \text{Tr}_{\rho
\sigma}^{(vertex)}$ is straightforward to perform. We look at the longitudinal
and transverse parts separately. We obtain,
\bea
&& P_L^{\rho\sigma} \text{Tr}_{\rho \sigma}^{(vertex)} = 8(1-x^2)
\left[
q_0^2(2p^2 +l(p_0 x+p y) - q_0 l(-2p^2 x+ p_0(l(1-x^2) + 2p y)) - 2l^2 p^2(1-y^2)\right] \nonumber\\
&& P_T^{\rho\sigma} \text{Tr}_{\rho \sigma}^{(vertex)} = 16\left[\right.q_0^2(p^2(1-y^2)+l(p_0 x-p y)) +q_0(l^2(p_0(1+x^2)-2 p x y)-2 l p(p_0 y - p x)) \nonumber \\
&&~~~~~~~~~~~~~~~~~~~~-l^2 p(4p_0 x y -p(1+x^2)(1+y^2))\left.\right]
\eea
where we have used $x:=l_0/l$ and set $P^2=0$ in anticipation of the fact that
the factor $\rho(P)$ in (\ref{vertex2}) will give $P^2=0$ when (\ref{rho}) is
used.

\subsection{Self-energy graph}

The expression for the self-energy diagrams is similar.  We obtain the
following expression for the diagram of Fig.~[\ref{figdiagrams}b]:
\begin{eqnarray}
\text{Im} \ {\Pi_R^{self (b)}}^{\mu}_{\mu}  &=&  - N_c C_F {e^2 g^2 \over 2}  \int
	\frac{d^4 P}{ ( 2\pi )^4} \left[ n_F (r_0) - n_F (p_0) \right] 
	\int \frac{d^4 L}{ ( 2 \pi )^4}\, \text{Tr}_{\rho \sigma}^{(self (b))}
n_B (l_0) {\cal R}^{\rho\sigma}  \nonumber
\\
&&\qquad ~~ \times \Bigl[D_R(P)-D_A(P)\Big]\ \Bigl[(D_R(R))^2 D_R(R+L) - (D_A(R))^2
D_A(R+L)\Bigr]
\label{self2-2}
\end{eqnarray}
where, as before, we have kept only the terms proportional to $n_B (l_0)$.
Rewriting using (\ref{convert}) we obtain,
\begin{eqnarray}
\text{Im} \ {\Pi_R^{self (b)}}^{\mu}_{\mu}  &=&  N_c C_F {e^2 g^2 }  \int
	\frac{d^4 P}{ ( 2\pi )^4} \left[ n_F (r_0) - n_F (p_0) \right]\;\rho(P) 
	\int \frac{d^4 L}{ ( 2 \pi )^4}\, \text{Tr}_{\rho \sigma}^{(self (b))}
n_B (l_0) {\cal R}^{\rho\sigma}  \nonumber
\\
&&\qquad\qquad\qquad ~~ \times \Bigl[ {\cal P}(R+L)~ {\cal P}(R)~ \rho(R) +
({\cal P}(R))^2 ~\frac{1}{2}\rho(R+L) \Bigr]
\label{self2}
\end{eqnarray}
As in the case of the vertex diagram, the first term in the square
bracket corresponds to the real cut, and the second term corresponds to the
virtual one.
The Dirac trace is: 
\bea
\text{Tr}_{\rho \sigma}^{self (b)} && = \text{Tr} \left[ \slP \ \gamma^{\mu} \
\slR \ \gamma_{\rho} \ ( \slR + \slL  )  \ \gamma_{\sigma} \ \slR \
\gamma_{\mu} \right] \nonumber \\ && = 8(g_{\rho\sigma}(R^2 (R\cdot P -L\cdot
P)+2 L\cdot R R\cdot P)+2(L_\rho+R_\rho)(R^2 P_\sigma -2 R\cdot P R_\sigma))
\label{tr_self}
\eea
The expression for the second self-energy diagram (Fig.~[\ref{figdiagrams}{\it
c}]) can be obtained in the same way and it is taken into account by 
multiplying eq.~(\ref{self2}) by a factor two. Contracting
$P_{T/L}^{\rho\sigma} \text{Tr}_{\rho \sigma}^{self(b)}$ we obtain,
\bea
&&P_L^{\rho\sigma} \text{Tr}_{\rho \sigma}^{self(b)} = 8(1-x^2) q_0^2(q_0 p_0+l(p_0 x-p y))  \nonumber\\
&&P_T^{\rho\sigma} \text{Tr}_{\rho \sigma}^{self(b)} =-16 q_0^2(p_0 q_0+(l(p_0 x+ p y) + p^2(1+y^2)) 
\eea
where we have set $P^2=0$ as before.

 The first step in the calculation of (\ref{vertex2}) and (\ref{self2}) is to
 perform the $p_0$ integration using (\ref{rho}). The choice $q_0>0$ means that
 only $n=1$ is kinematically allowed.  The second step is to use the delta
 functions contained in the spectral functions $\rho(R+L)$ and $\rho(R)$ to
 perform the $p$ integrations in the real and virtual terms respectively. The
 structure of the real cut is the same for both vertex and self energy
 diagrams. Making use of (\ref{rho}) we have,
 \bea
 \rho(R+L) =  \frac{\pi}{ l_0+  q_0 + l y}\delta(p-p_{{\rm real}})\,;~~~p_{{\rm real}} = \frac{(l_0+q_0)^2 - l^2}{2( l_0 +  q_0 + l y)}
 \label{deltap}
 \eea
Similarly, we attack the virtual cut by using (\ref{rho}) to obtain
(remember that $r_0 = q_0+p_0$ and $r=p$ in static photon kinematics),
\bea
\rho(R) =  \frac{\pi}{q_0}\delta(p-p_{\rm virtual})\,;~~~p_{{\rm virtual}} = \frac{q_0}{2} 
\eea
This expression can be used directly to do the $p$ integral in the
virtual part of the vertex diagram. The structure of the virtual part of the
self energy diagram is more complicated: there is a singularity because of the
presence of the factor ${\cal P}(R)~\rho(R)$. This singularity can be
regulated in the standard way \cite{LeBellacBook}. We write
\bea
\left(\frac{1}{x+i\epsilon}\right)^2 = -\frac{d}{d\,x}\left(\frac{1}{x+i\epsilon}\right).
\eea
Taking the imaginary part of this equation gives,
\bea
{\cal P}\left(\frac{1}{x}\right) \,\delta(x) = -\frac{1}{2}\frac{d}{d\,x}\delta(x)
\eea
Using this expression we obtain, 
\bea
{\cal P}(R)~\rho(R) = \frac{\pi }{4 q_0^2}\,\frac{d}{d\,\,p} \delta(p-p_{\rm virtual})
\eea
and perform an integration by parts to remove the derivative from the delta
function.

Once the variables $p_0$ and $p$ have been integrated over, we can simplify
the result by performing an expansion in $l_0/q_0$ and $l/q_0$. However, as
shown below, one finds  that the leading order term cancels between
the real and virtual contributions to each diagram, and thus we must expand to
next-to-leading order. The fact that  $p_{\rm real}$ is a function of $l_0$,
$l$ and $q_0$ means that the contributions from the real cuts will involve an
expansion of the $(n_F (r_0) - n_F (p_0))$ thermal factors. We use the
notation:
\begin{equation}
n_F = \frac{1}{e^{\beta q_0/2}+1} \ ;\quad n_F' = 0\ ;\quad n_F'' =
n_F(1-n_F)(1-2n_F)
\label{eq:nf}
\end{equation}
After the expansion has been performed, the resulting expression can be
integrated over $y$ by hand. We look at longitudinal and transverse pieces separately.

\section{Preliminary Results}
\label{Results}

We list below the results for the trace of the imaginary part of the retarded
polarization tensor. To simplify the notation, a common factor has been
extracted from each term:
\bea
\text{FAC} = -\frac{e^2 g^2}{16\pi^4}N_c~C_F \int dl\,l \int dx\, n_B(x \,l)
\rho_{L/T}(lx,l) 
\label{factor}
\eea
In addition, we have used the notation:
\bea
\text{LOG} = {\rm ln} \left| \frac{x+1}{x-1}\right|~;~~~~x=\frac{l_0}{l} \nonumber
\end{eqnarray}
We obtain:
\begin{eqnarray}
{\rm Vertex}_L^{\rm virtual} & = & (1-2 n_F)~\frac{1-x^2}{x}~\left(q_0^2
\text{LOG}-2l^2 (2x +(1-x^2)\text{LOG} \right) 
\label{VVL}\\
{\rm Vertex}_L^{\rm real} & = & -(1-2 n_F)~\frac{1-x^2}{x}~\left[q_0^2
\text{LOG}-2l^2 (x +(1-x^2)\text{LOG}\right]\nonumber \\ &&~~~~ -\frac{1}{2}
l^2 \frac{q_0^2}{T^2}  (1-x^2)(1-x\text{LOG}) n_F''
\label{VRL}\\
{\rm Vertex}_L & = &-l^2(1-x^2)\left(2(1-2 n_F)+ \frac{1}{2}\frac{q_0^2}{T^2}
(1- x\text{LOG})n_F''\right)
\label{VL}\\
&& \nonumber \\
&& \nonumber \\
&& \nonumber \\
{\rm Self-Energy}_L^{\rm virtual}& = &2(1-2 n_F) ~q_0(q_0 + 2 l x) 
\label{SEVL}\\
{\rm Self-Energy}_L^{\rm real} & =  &-2(1-2 n_F)~\left[q_0^2+2l q_0 x - l^2
(1-x^2) \right]-\frac{1}{2}l^2 \frac{q_0^2}{T^2}(1-x^2)(1+ \frac{2 x l}{q_0})
n_F''
\label{SERL}\\
{\rm Self-Energy}_L & = &l^2(1-x^2)\left(2(1-2 n_F)-
\frac{1}{2}\frac{q_0^2}{T^2}(1+ \frac{2 x l}{q_0})n_F''\right)
\label{SEL}
\eea
\\
\bea
{\rm LONG} \ &=& \ -\frac{1}{2} l^2 \frac{q_0^2}{T^2}(1-x^2)(2- x
\text{LOG}+\frac{2 l x}{q_0})n_F''
\label{LONG}\\
&& \nonumber \\
&& \nonumber \\
&& \nonumber \\
{\rm Vertex}_T^{\rm virtual}& = &(1-2n_F)~\frac{1}{x}~(q_0^2-2l^2)(2x
+(1-x^2)\text{LOG}) \\
{\rm Vertex}_T^{\rm real}& = &-(1-2n_F) ~\frac{1}{x}~[q_0^2(2x
+(1-x^2)\text{LOG})-2l^2(1-x^2)\text{LOG}] \nonumber \\
&&~~~~- \frac{q_0^2}{T^2} l^2 \left( \frac{1}{3}(1-3x^2) - \frac{1}{2}
x(1-x^2)\text{LOG}\right) n_F'' \\
{\rm Vertex}_T & = &-l^2\left( 4(1-2n_F)+ \frac{q_0^2}{T^2}
\left(\frac{1}{3}(1-3x^2)-\frac{1}{2}x(1-x^2)\text{LOG}\right)n_F''\right)\\
&& \nonumber \\
&& \nonumber \\
&& \nonumber \\
{\rm Self-Energy}_T^{virtual} & = & -2(1-2n_F)[q_0^2(2-x \text{LOG})+l^2 x
(2x-(1+x^2)\text{LOG}] \\
{\rm Self-Energy}_T^{real} & = &2(1-2n_F)\left[(2-x
\text{LOG})(q_0^2+l^2(1+x^2))\right]+\frac{1}{3}\frac{q_0^2}{T^2} l^2 n_F'' \\
{\rm Self-Energy}_T & = &l^2\left(4(1-2n_F)+\frac{1}{3} \frac{q_0^2}{T^2} 
n_F''\right)\\
&& \nonumber \\
&& \nonumber \\
&& \nonumber \\
\text{TRANS} \ &=& \ \frac{1}{2}l^2 \frac{q_0^2}{T^2}x(2x+(1-x^2)\text{LOG})n_F''
\label{TRANS}
\eea
Before proceeding, we make the following observations about these results:

\noindent [1]: 
In each of the four cases: 
(vertex/self-energy)$\times$(longitudinal/transverse), the leading order term
in the $l/q_0$ expansion cancels between real and virtual contributions and the
final result is of order $l^2$.

\noindent [2]: For both the longitudinal and transverse pieces the term
proportional to $(1-2n_F)$ cancels between the self energy and vertex
contributions.

\noindent [3]:  The longitudinal propagator has the form $\Delta_L(L)\sim
1/(1-x^2)$, as can be seen from (\ref{DeltaTL}) and (\ref{PiTL}). Consequently,
the terms (\ref{SEVL}) and (\ref{SERL}) exhibit co-linear singularities in the
form of a factor $1/(1-x^2)$ which diverges when $x\rightarrow \pm 1$. When the
two expressions are combined however, the sum (\ref{SEL}) is well behaved.

\vspace*{.5cm}

\noindent As a check on these results, we sketch in the appendix an alternative derivation of the
transverse gluon contribution.

\section{$l_0$ and $l$ integrals}
\label{Int}

In this section we perform the two remaining integrals over $l_0$ and $l$.
From (\ref{factor}), (\ref{LONG}) and (\ref{TRANS}) and using the approximation $n_B (lx) \sim T/ (lx)$ we have,
\bea
&& LONG = -\frac{e^2 g^2}{(2\pi)^4}N_c ~ C_F ~\frac{q_0^2}{T}~ n_F'' ~ I_L \nonumber \\
&& TRANS = -\frac{e^2 g^2}{(2\pi)^4}N_c ~C_F ~\frac{q_0^2}{T}~ n_F'' ~ I_T \nonumber 
\eea
where,
\bea
\label{IL}
&&  I_{L}  = \int^\infty_{-\infty}\frac{dx}{x} \int ^\infty_0
dl\,l^2\,\rho_L(lx,l)\, \left[(1-x^2)\,\frac{x}{2}\ln \left|
\frac{x+1}{x-1}\right|-1 \right] \\ 
\label{IT}
&& I_{T}  = \int^\infty_{-\infty} \frac{dx}{x}  \int_0^{\infty} dl\; l^2 \:
\rho_T (lx,l) \, \left[ x^2 + (1-x^2) \frac{x}{2} \ln \left|
\frac{x+1}{x-1}\right| \right] 
\eea
It is interesting to remark that the terms in square brackets in the integrands are proportional to ${\rm Re} \Pi_L$ and ${\rm Re} \Pi_T$ respectively.
 
\subsection{Longitudinal Integral}

We consider first the longitudinal integral~\eqref{IL}. We divide the integral into two pieces, one with and one without the logarithm:
\bea
&&  I^a_{L}  = -\int^\infty_{-\infty}\frac{dx}{x} \int ^\infty_0 dl\,l^2\,\rho_L(l\,x,l)\,(1-x^2)  \\
\label{longa}
&&  I^b_{L}  = \int^\infty_{-\infty}\frac{dx}{x} \int ^\infty_0 dl\,l^2\,\rho_L(l\,x,l)\,(1-x^2) ~\frac{x}{2}\ln \left| \frac{x+1}{x-1}\right| 
\label{longb}
\eea 
The calculation of the first piece is standard \cite{Aurenche:1996sh}. One obtains
\bea
I_L^a = \sqrt{3}\pi^2 m_g 
\label{Ia}
\eea
To calculate $I_L^b$ we use the  method proposed in~\cite{LeBellac:qg}.  We use an integral representation for the logarithm and write,
\begin{equation}\label{idlog1}
\ln \left| \frac{x+1}{x-1} \right| = 
   \int_{-1}^{1} dx' {\cal P}\left( \frac{1}{x-x'}
  \right)  
\end{equation}
where the principal part is defined as 
\bea
{\cal P}\left(\frac{1}{x}\right) = \frac{1}{2}\left(\frac{1}{x+i\epsilon} + \frac{1}{x-i\epsilon}\right)
\nonumber
\eea
The first step is to perform the integral 
\bea
\int d x {1 \over 2} (1-x^2)\ \rho_L(l x, l)\ {\cal P}\left( \frac{1}{x-x'}
  \right)  
\eea  
via the standard dispersion relation (see \cite{LeBellacBook}) keeping track
of the contribution of the large circle at infinity. 
We obtain 
 \bea
  I_L^b = -\pi {\rm Re} \int^1_{-1} dx' \int_0^\infty dl
\left(1-\frac{l^2}{l^2-3m_g^2 Q_1(x')}\right)\,;~~Q_1(x') = \frac{x}{2}\ln 
\left(\frac{x'+1}{x'-1}\right)-1,
  \eea
and performing the integral over $l$ we have,
  \bea
  I_b = \frac{\sqrt{3}}{2} \pi m_g \int ^1_{-1} dx' {\rm Re} \sqrt{Q_1(x')}\left( {\rm ln} (\sqrt{Q_1(x')})-{\rm ln}(-\sqrt{Q_1(x')})\right)
  \eea
  
  \noindent Cutting the complex plane along the negative real
axis for both the square root function and the logarithm we find,
  \bea
  \left( {\rm ln} (\sqrt{Q_1(x')})-{\rm ln}(-\sqrt{Q_1(x')})\right) = -i\pi \,{\rm sgn}(x')
  \nonumber
  \eea
  which gives,
  \bea
  I_b = \frac{\sqrt{3}}{2} \pi^2 m_g {\rm Im}\int ^1_{-1} dx'{\rm sgn}(x')\sqrt{Q_1(x')}
  \eea
  We do the $x'$ integral by writing $f(z) = \sqrt{Q_1(z)}$ and choosing the contour shown in Fig. (\ref{figdiagrams3}).
We obtain,
  \bea I_b = -\pi^3 m_g /2 \label{Ib}
  \eea
  
  \begin{figure}[ht]
   \centering
\includegraphics[width=0.15\textwidth]{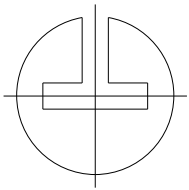}
\caption{Contour in the complex $x'$ plane}
\label{figdiagrams3}
\end{figure}
\noindent Combining (\ref{Ia}) and (\ref{Ib}) we have,
  \bea
  I_L = \pi^2 m_g \left(\sqrt{3}-\frac{\pi}{2}\right)
  \eea
 
 \subsection{Transverse Integral}

In this section we consider the transverse integral \eqref{IT}. The
$x$-integral can be performed in a way similar to the one above, but the result
exhibits an ultraviolet divergence in the $l$-integral. To understand this
result, we separate the gluon spectral function into the pole piece and the cut
piece (see \cite{LeBellacBook}). 

The pole piece carries the ultraviolet divergence. This divergence is not
physical: it is a consequence of performing an infrared expansion on an
integral that is dominated by the ultraviolet part of the phase space (the
expansion in $l/q_0$ used above is not valid in this case). But the pole part
is just the contribution calculated in ~\cite{Altherr:1989jc} (up to higher
order terms since the transverse gluon has a small thermal mass), the result
of which is given in eqs.~(\ref{eq:2-bareloop}),~(\ref{eq:2-bareloop2}).

The cut part of the spectral function does not produce a divergence.  We have,
\begin{equation}
I_T^{\rm cut} = \int_{-1}^{1} \frac{dx}{x} \int_0^{\infty} dl \,\,l^2 \beta_T (x,l)
\left[ x^2 + (1-x^2) \frac{x}{2} \ln \left| \frac{x+1}{x-1}
\right| \right]
\end{equation}
where
\bea
\beta_T(x,l) = \frac{-2\,\,{\rm Im}\,\,\Pi_T}{(l^2(x^2-1)-{\rm Re}\,\,\Pi_T)^2
+ ({\rm Im}\,\,\Pi_T)^2}
\nonumber
\eea
and $\Pi_T$ is given in (\ref{PiT}).  Since the integration range does not
extend from positive to negative infinity we cannot use a sum-rule-like trick
to perform the integral. We proceed as follows. The $\int dl$ integral can be done by hand. The only mass scale in the integrand is $m_g$ and a simple dimensional analysis shows that the $\int dx$ must be of order $m_g \sim g T$.   The remaining integral is done numerically to obtain,
\bea
I_T^{\rm cut} = -6.67~ m_g
\eea
which is of the same order as the longitudinal contribution.

\section{Conclusions}
\label{Conclusions}

In this note we have computed the lepton pair production rate by a static
heavy photon at two-loop order~(Fig.~[\ref{figdiagrams}]) and focused on the
infrared region for the exchanged gluon. The result is free of infrared and
collinear divergences, in agreement with a previous study in the case where
the external particle was a scalar~\cite{LeBellac:qg}. As
in~\cite{LeBellac:qg}, we find the contribution of the longitudinal gluon to
be of order~$e^2 g^3$. However, our results differ for the transverse sector.
The authors of~\cite{LeBellac:qg} find that the contribution of the
transverse gluon is of order~$e^2 g^4$. Our results are as follows: 

\noindent [1] The infrared approximation used in this computation is valid
for the ``cut'' part ($x^2<1$) of the gluon spectral function. The
contribution for ``cut'' part is of order~$e^2 g^3$.

\noindent [2] The infrared region of the gluon phase space does not give the
dominant contribution to the pole piece of the gluon spectral function. The
pole piece is dominated by large momenta. This means that thermal effects can
be neglected and we can use the result obtained with a bare gluon propagator
which was calculated previously~\cite{Altherr:1989jc}.

The physical picture of the calculated contributions is the following. The
one-loop contribution is the annihilation of a quark-antiquark pair, each
parton carrying half of the energy of the static photon, namely $q_0/2$.
The dominant two-loop contribution describes the scattering of one of the
annihilating quark in the medium with an exchange of small momentum  of order
$l \sim m_g \sim g T$. Interestingly, these processes are similar to those
which dominate the production of the small mass, high momentum (of ${\cal
O}(T)$ or higher) lepton pairs, {\em i.e.} the other extreme kinematical
domain~\cite{Aurenche:2002}. In the static case, higher order loop diagrams
may modify the result, as was found to be the case for energetic photons with a small or vanishing invariant mass~\cite{Arnold:2001,Aurenche:2002-1}.

We summarize below the result for ~$\text{Im} {\Pi_R}^{\mu}_{\mu}$ up to
order~$g^3$ in perturbation theory~:
\begin{equation}\label{Res1}
\text{Im} {\Pi_R}^{\mu}_{\mu} =  \alpha \ N_c \ q_0^2 \left[ 1-2 \, n_F
\right] \left[ 1 +  C_F  {g^2 \over 4 \pi^2} \left( {3 \over 4}\ + ~ 1.62 ~
n_F (1-n_F) {m_g \over T} \right) \right]
\end{equation}

The complete result~\eqref{Res1} shows that the processes considered in this
paper can be relatively large when $q_0/T$ is not too large. This suggests that
these processes may be relevant for lepton pair production from static heavy
photons at temperatures where we expect a quark-gluon plasma to be
formed~\cite{Ullrich:wt,Lenkeit:1999xu,Adamova:2002kf}.


\section{Appendix}
\label{appendix}

As a check of the above results we have derived the transverse gluon
contribution using an alternative method. The $\int dp_0$ integration is done using
Cauchy's theorem in the complex $p_0$ plane. The $\int dy$ integration is performed second, and then the imaginary part is taken. The $\int dp$ integration is done last. In addition, vertex and self-energy matrix elements are
combined before carrying out the integral so that a partial cancellation of
terms occurs before integration. We start from (\ref{vertex2-2}) and (\ref{self2-2}).
In the $p_0$ complex plane, the integrand has poles above and below the real
axis from the propagators, as well as poles on the imaginary axis from the
statistical factors  $n_F (r_0)$ and $n_F (p_0)$. First we note that when we close the contour in the
upper-half or the lower-half $p_0$ plane, the contribution from the poles on the
imaginary axis drop out because the difference between the retarded and the
advanced propagators vanishes. Second, we use the fact that terms containing only retarded or advanced
propagators do not contribute. The product of propagators can thus be simplified so  that we obtain: 
\begin{eqnarray}
\text{Im} \ {\Pi_R^{self (b)}}^{\mu}_{\mu}  &=&   N_c C_F e^2 g^2  {\rm Re}~ \int
	\frac{d^4 P}{ ( 2\pi )^4} \left[ n_F (r_0) - n_F (p_0) \right] 
	\int \frac{d^4 L}{ ( 2 \pi )^4}\, \text{Tr}_{\rho \sigma}^{(self (b))}
n_B (l_0) {\cal R}^{\rho\sigma}  \nonumber
\\
&&\qquad ~~~~~~~~~~~~~ \times \Bigl[D_A(P)(D_R(R))^2 D_R(R+L) \Bigr]
\label{self2-3}
\end{eqnarray}
\bea
\text{Im} {\Pi_R^{(vertex)}}^{\mu}_{\mu} &=&   N_c C_F e^2 g^2  {\rm Re}~\int 
\frac{d^4 P}{ ( 2\pi )^4} \left[ n_F (r_0) - n_F (p_0) \right]  \int \frac{d^4
L}{ ( 2 \pi )^4}\, \text{Tr}_{\rho \sigma}^{(vertex)} n_B (l_0) {\cal
R}^{\rho\sigma}  
\nonumber \\
&&\qquad\qquad\qquad\times \ \Bigl[D_R(P) D_R(P+L)D_A(R)D_A(R+L)\Bigr] 
 \label{vertex2-3}
\eea
In the expressions above it is understood that the poles in the statistical weights should be ignored when doing the $p_0$ integration in the complex plane.
Only the poles
from the propagators contribute.\\

We can now add vertex and self-energy contributions so that the full expresion
of the trace on the quark loop multiplied by the appropriate propagators is:
\bea
&& \rm Tr_{\rho \sigma} \cdot {\rm propagators} = \nonumber \\
&& \qquad
- 4\ \left[
{2 L^2 (R^\rho R^\sigma + P^\rho P^\sigma) - 4 Q^2 R^\rho P^\sigma +
g^{\rho\sigma} L^2 ( Q^2 + L^2 -P^2 -R^2 -(P+L)^2 -(R+L)^2) 
\over P^2_R R^2_A (P+L)^2_R (R+L)^2_A} \right.\nonumber \\
&&  \qquad 
+\ Q^2 \left({-4 P^\rho P^\sigma + g^{\rho\sigma} ((P+L)^2  -L^2) 		
		\over (P^2_R)^2 R^2_A (P+L)^2_R} +
	{- 4 R^\rho R^\sigma + g^{\rho\sigma} ((R+L)^2  -L^2)
	\over P^2_R (R^2_A)^2 (R+L)^2_A} \right) \label{eq:trace}\\
&&  \qquad \left.
+\ g^{\rho\sigma} \left({1 \over P^2_R}-{1 \over (P+L)^2_R} \right)
		 \left({1 \over R^2_A}-{1 \over (R+L)^2_A} \right)
+\ g^{\rho\sigma}\left({2 Q L \over P^2_R R^2_A (P+L)^2_R}	
			- {2 Q L \over P^2_R R^2_A (R+L)^2_A} \right) \right] 
\nonumber
\eea
where we have used the notation $P_R^2 = (p_0+i\epsilon)^2-p^2$, etc. 
As discussed in \cite{Aurenche:2002-1} (see Eq.(3.12) of this reference), this expression
already exhibits some cancellations due to gauge invariance. Further simplifications are possible in our case since we are working in the limit of large $q_0^2=Q^2$ and small $\{l^2_0,~l^2,~L^2\}$.  At this point we consider only transverse gluons (longitudinal gluons can be treated using the same method) and identify the terms that contribute to (\ref{eq:trace}) to leading order. It has been checked
explicitly that all other terms are subdominant. We need to calculate: 
\bea
&& T_1 = e^2 g^2  N_c C_F  {\rm Re} \int  \frac{d^4 P}{ ( 2\pi )^4} 
\left[ n_F (r_0) - n_F (p_0) \right]  
\int \frac{d^4 L}{(2 \pi)^4}\ n_B(l_0) \rho_T(L) 
{16\ Q^2\ R^\rho P^\sigma  P_T^{\rho \sigma}(L) \over P^2_R R^2_A (P+L)^2_R
(R+L)^2_A} \nonumber \\
&& T_2 = e^2 g^2  N_c C_F  {\rm Re} \int  \frac{d^4 P}{ ( 2\pi )^4} 
\left[ n_F (r_0) - n_F (p_0) \right]  
\int \frac{d^4 L}{(2 \pi)^4}\ n_B(l_0) \rho_T(L) 
{16 \ Q^2\ R^\rho R^\sigma  P_T^{\rho \sigma}(L) \over P^2_R (R^2_A)^2 
(R+L)^2_A} \label{bigterms}
\eea
and an equation similar to $T_2$ with $P$ and $R$ interchanged.
We discuss below the strategy for calculating the necessary integrals. Consider, for example, a term of the form:
\begin{equation}
{\rm Int} = {\rm Re} \int p^2\,dp\ dy\ dp_0\ {1 \over P^2_R (P+L)^2_R  R^2_A (R+L)^2_A}
\label{int1}
\end{equation}
where the integrals over $l_0$ and $l$ have been supressed (see (\ref{phasespace})). These integrals must be calculated following the method of Sec.~\ref{Int} and are not discussed in this appendix.
We close the $p_0$ contour in the lower half-plane. There are contributions from the poles of the factors $P_R^2$ and $(P+L)_R^2$.  
In the small $\{l_0,~l\}$ limit the contributions from the poles of the factor $(P+L)_R^2$ will have the same form as the contribution from the poles of $P_R^2$. Thus we will consider only the poles $p_0 = \pm p - i \epsilon$ from the factor $P_R^2$, and multiply the result by a factor of two.
This procedure gives rise to an  integral of the form,
\bea
\label{PAdelta}
{\rm Re} \int dp_0 { F(p_0) \over P^2_R} &=&
\sum_{n=\pm} {\rm Re}\left( {i \pi \over p} F(n p - i \epsilon)\right) \nonumber \\
&=& - {\pi \over p}\ {\rm Im}\ F(- p - i \epsilon),
\eea
where we have neglected, in the last line, the pole which will not lead to an
imaginary part in the function $F$. Using (\ref{PAdelta}) we obtain,
\bea
&&{1 \over R^2_A} = {1 \over q_0 (q_0 - 2p - i \epsilon)},\qquad 
{1 \over (P+L)^2_R} = {1 \over c - b y}, \qquad
{1 \over (R+L)^2_A} = {1 \over  a - b y} 
\label{denoms}
\eea
where
\bea 
a - b = (q_0 + l_0 + l)\ (q_0 - 2 p + l_0 - l) - i \epsilon , \qquad &&
a + b = (q_0 + l_0 - l)\ (q_0 - 2 p + l_0 + l) - i \epsilon \nonumber \\
c - b = (l_0 + l)\ (- 2 p + l_0 - l) ,\qquad\quad\qquad\qquad \  &&
c + b =    (l_0 - l)\ (- 2 p + l_0 + l).
\label{abc}
\eea
We note that when the retarded
propagator $1/(P+L)^2_R$ is evaluated at the pole of the retarded function $1/P^2_R$, the $i \epsilon$ piece cancels exactly and thus the variables $b-c$ and $b+c$ in (\ref{abc}) do not contain an $i\epsilon$ prescription. Carrying out the angular integration on $y$, one obtains:
\bea
\int^1_{-1} dy\ {1 \over (a - b y)} {1 \over (c - b y)} = {1 \over 2 q_0 p l}
{1 \over (q_0 - 2 p + 2 l_0 - i \epsilon)}\ \left(\ln \left({a-b \over
a+b}\right) - \ln \left({c-b \over c+b}\right) \right).
\label{eq:int-y}
\eea
Combining (\ref{int1}), (\ref{PAdelta}), (\ref{denoms}) and (\ref{eq:int-y}) we have,
\bea
&& {\rm Int} \label{int2} \\
&& ~~= -{ 2} \pi \int dp\, p\, {\rm Im}\left[{1 \over 2 q_0 p l}{1\over q_0(q_0-2p-i\epsilon)}
{1 \over (q_0 - 2 p + 2 l_0 - i \epsilon)}\ \left(\ln \left({a-b \over
a+b}\right) - \ln \left({c-b \over c+b}\right) \right)\right].\nonumber 
\eea
where
\bea
{1 \over (q_0 - 2p - i \epsilon)(q_0 - 2 p + 2 l_0 - i \epsilon)} &=& 
{1 \over 2 l _0} \left({\cal P}\left({1 \over q_0 - 2p}\right)  - {\cal P} 
\left({1 \over q_0 - 2 p + 2 l_0}\right) \right) \nonumber \\
&&  
+{i \pi \over 2 l_0}\ \left(\delta(q_0 - 2p) - \delta(q_0 - 2 p+2 l_0)\right).
\label{eq:pole}
\eea
There is also an imaginary piece to the first logarithm in (\ref{int2}). We use the relation $\ln (x - i\epsilon)
=  \ln(|x|) - i \pi \theta(-x)$ to separate real and imaginary pieces and obtain,
\bea 
\int^\infty_0 dp \ln \left({a-b \over a+b}\right) =
\int^\infty_0 dp \left( \ln\left( {q_0+l_0+l \over q_0+l_0-l}\right) + 
\ln\left( {q_0-2p+l_0-l \over q_0-2p+l_0+l}\right) \right) 
 -i \pi \int^{(q_0+l_0+l)/2}_{(q_0+l_0-l/2} dp
\label{eq:log}
\eea
To complete the evaluation of (\ref{int2}) we take the product of (\ref{eq:pole}) and 
(\ref{eq:log}) and extract the imaginary part. There are two contributions: one contains the imaginary part of (\ref{eq:pole}) which contains a delta function that allows us to do the $\int p$ integral, and the second contains the imaginary part of (\ref{eq:log}). This factor contains an integral in $p$ phase-space that 
is restricted to a small interval around $q_0/2$ which can be calculated by expanding around the value $p=q_0/2$.   This procedure leads to the appearance of a factor $n''_F$ as defined in (\ref{eq:nf}). The $\int dp$ integration can then easily be done. The calculation of terms like $T_2$ is carried out in the same way, the double pole being handled via an integration by parts on the momentum variable after taking the residue of the single pole via the contour integration. Finally we obtain: 
\bea
T_1
 = {e^2 g^2 \over (2 \pi)^4} N_c C_F {q^2_0 \over T} n''_F \ 
\int {dx \over x} \int dl\ l^2 \rho_T(x,l)
\left[ x^2 - {1\over 3} + (1-x^2) {x \over 2} \ln\left(1+x \over 1-x\right) 
\right]
\label{eq:vertex2-3}
\eea
\bea
T_2 = {e^2 g^2 \over (2 \pi)^4} N_c C_F {q^2_0 \over T} n''_F \ 
\int {dx \over x} \int dl\ l^2 \rho_T(x,l) {1\over 6}.
\qquad\qquad\qquad\qquad\qquad\qquad\  
\label{eq:self2-3}
\eea
Adding (\ref{eq:vertex2-3}) and twice (\ref{eq:self2-3}) gives the full
transverse contribution as calculated in (\ref{IT}).


\end{document}